# Application of compact TiO$_2$ layer fabricated by pulsed laser deposition in organometal trihalide perovskite solar cells

Hao Zhang [1,&], Hong Wang [1,&], Meiyang Ma [1], Yu Wu [1], Shuai Dong [1,*], and Qingyu Xu [1,2,*]

[1]School of Physics, Southeast University, Nanjing 211189, China

[2]National Laboratory of Solid State Microstructures, Nanjing University, Nanjing 210093 China

Abstract:

Organometal trihalide perovskite solar cells have been rapidly developed and attracted much attention in recent years due to their high photoelectric conversion efficiency and low cost. Pulsed laser deposition (PLD) is a widely adopted technology which is used in the preparation of thin films, especially oxide thin films. With this technology, the thickness and composition of films can be conveniently and accurately controlled. In the structure of perovskite solar cells, TiO$_2$ layer working as the n-type semiconductor is used to block holes and transport electrons into electrode, which is crucial for the performance of whole devices. We introduced the PLD technique into preparation of TiO$_2$ layer. In comparison with common spin coating method, TiO$_2$ layer prepared by this technique is ultrathin and more compact. Compact TiO$_2$ (c-TiO$_2$) layers with optimized thickness of 32 nm have been prepared by the PLD method and the highest efficiency of 13.95 % for the MAPbI$_3$-based solar cell devices has been achieved.

&These authors contribute equally to this work.

*Corresponding authors: sdong@seu.edu.cn (S.D.); xuqingyu@seu.edu.cn (Q.X.)

1. Introduction

Due to serious environmental issue produced by traditional fossil fuels which are also unable to meet fast increasing energy requirement of modern society and industry, renewable clean resources are essential for the development of human society.[1] Solar energy as the most abundant energy resource can meet the demand of human if they are utilized effectively and can be mainly achieved by solar photovoltaic cells currently.[2] Among all sorts of solar cells, organic-inorganic halide perovskite solar cells (PVSCs) remarkably attracted great attention and developed very fast in the past few years due to its rapid increase of power conversion efficiency (PCE) and low fabrication cost.[3,4] The organic-inorganic halide perovskite materials with formula ABX$_3$ (A cations are organic macromolecule, B cations are metal elements and C anions are halogen elements) have cubic crystal structure [5,6] and possess many merits such as long diffusion lengths,[7] appropriate direct bandgap,[1] high absorption coefficient[8] and low trap state density[9]. Currently, the highest PCE of PVSCs has reached 22.1% and the theoretical upper limit of PCE is 31%.[10,11] The electron transport layer in PVSCs plays a crucial role because its function is to prevent carriers from directly contacting the conductive substrate which results in serious charge carrier recombination.

Titanium dioxide ($TiO_2$) is one of the commonest and excellent materials applied in photovoltaic devices as the electron transport layer (ETL) since it is a wide-bandgap semiconductor and fairly transparent.[12] $TiO_2$ has three crystal structure: rutile, anatase and brookite,[13] and the $TiO_2$ layers used in PVSCs with single-phase anatase crystal have excellent optical property which are generally prepared by chemical method at temperature higher than 500 °C.[14,15] Pulsed laser deposition (PLD) is a well-known technology, which is widely used to fabricate oxide thin films owing to its advantages: the simple system operation, a wide range of deposition condition, rich choice of materials, and so on.[16] Most importantly, with PLD, high-quality thin films with high reproducibility can be prepared by the adjustment of deposition temperature, laser frequency, the gas introduced and partial pressure, selected substrate and other conditions.[17]

As we all know, solution methods which are widely applied in PVSCs like spin coating are unable to deposit uniform thin films of large area so that they cannot be used in preparation of massive-area solar cells.[18] So, new technology is essential to be adopted in preparation of c-$TiO_2$ layers. Now, the common technique applied in preparation of large-size perovskite solar cells is spray pyrolysis.[19,20] However, preparation of c-$TiO_2$ can be easily influenced by different operator and surrounding conditions. In comparison with the spray pyrolysis, the preparation of c-$TiO_2$ layers can be controlled easily and accurately by PLD. We can adjust the thickness and control the quantity by controlling the deposition parameters, such as pulse number, laser energy density, vacuum pressure, gas, annealing temperature and so on. In this work, we introduce PLD to prepare c-TiO$_2$ layers in CH$_3$NH$_3$PbI$_3$-based PVSCs.[21] We explore the optimum preparing conditions of c-TiO$_2$ layer prepared by PLD, which is more uniform and thinner with optimum thickness of 32 nm than that of c-TiO$_2$ layers prepared by spin coating whose optimum thickness is beyond 60 nm.[22] A thinner c-TiO$_2$ layer can effectively transmit light and reduce series resistance of devices.[23] In addition, c-TiO$_2$ layers doped with other elements can be fabricated more easily by PLD in comparison with other methods. Finally, a photoelectric conversion efficiency of 13.95% with short-circuit current ($J_{sc}$) of 21.83 mA/cm$^2$, open-circuit voltage ($V_{oc}$) of 1.0 V, and fill factor (FF) of 63.98% were achieved.

## 2. Experimental details

Firstly, we chose F-doped SnO$_2$ (FTO) glass as substrate which was cut into the size of 2 cm×2 cm. Then we cleaned the substrates with detergent water, deionized water and absolute ethyl alcohol by ultrasonic. The following step was the preparation of TiO$_2$ layer by PLD. We prepared the target with pure TiO$_2$ powder and annealed it at 800 °C for 4 hours. Subsequently, we put the cleaned FTO glass substrate and target into the chamber with base pressure of less than 10$^{-3}$ Pa. The wavelength of the excimer laser was 248 nm and the laser energy was set to 100 mJ. The target to substrate distance was 5 cm. We set pulse repetition rate to be 5 Hz at room temperature and the film thickness was controlled by pulse number. The substrate was annealed at 500 °C for 1 hour after deposition. Then, a thickness of around 200 nm mesoporous TiO$_2$ (m-TiO$_2$) layer was spin coated with diluent (m-TiO$_2$ paste : ethyl

alcohol = 1 : 4) at 4000 rpm for 30 seconds and then annealed at 500 °C for 60 minutes at the heating rate of 5 °C.

Next to the preparation of the m-TiO$_2$ layer, we synthesized CH$_3$NH$_3$PbI$_3$ (MAPbI$_3$) solution by mixing 0.795 g MAI, 2.305 g PbI$_2$, 3 g DMF and 400 μL DMSO together and magnetic stirring for 4 hours. Then, we spin coated the MAPbI$_3$ solution on the m-TiO$_2$ layer at the rate of 4000 rpm for 30 seconds and 1 ml diethyl ether was dropped in the process of spin coating in order to wash away DMF. When the process of spin coating was finished, the substrate was heated at 60 °C for 1 minute and at 100 °C for 2 minutes successively. Afterwards, a hole transport layer (HTL) with thinness of about 250 nm was fabricated by spin coating the solution mixed by 288 μL TBP, 175 μL Li-TFSI solution (520 mg Li-TSFI in 1 ml acetonitrile), 290 μL Co-TFSI solution (300 mg Co-TSFI in 1 ml acetonitrile) and 20 mL chlorobenzene together on the MAPbI$_3$ layer at the rate of 5000 rpm for 30 seconds. After one night, a 100 nm thick Ag layer was deposited on the HTL as back electrode by thermal evaporation at the rate of 0.2-0.3 Å/s in the chamber under the pressure of $10^{-4}$ Pa.

XRD patterns were measured by an X-Ray diffractometer (Rigaku Smartlab3) with Cu Kα radiation to analyze the structure of the thin film. Morphology images of thin films were obtained by a scanning electron microscope (SEM, FEI Inspect F50). Energy dispersive spectrometer (EDS) was measured with OXFORD instruments attached on the SEM. We measured light absorption spectroscopy with ultraviolet-visible spectrophotometer (HITACHI U-3900). The steady-state photoluminescence (PL) spectra were obtained by Raman spectrometer made by Horiba Jobin Yvon using a laser of 532 nm wavelength to investigate the band gap of samples. Atomic force microscopy (AFM) images were obtained by a BioScope ResolveTM. Current-voltage (I-V) curves were measured by a Keithley 2400 under Newport Oriel 91.192 simulated illumination (AM1.5, 100 mw/cm$^2$). Electrochemical impedance spectroscopy (EIS) was measured by the electrochemical workstation (Chenhua Instruments CHI660b). Transient-state PL spectrum was obtained by the fluorescence spectrophotometer (Edinburgh FLS980). Incident photon-to-current conversion efficiencies (IPCE) were measured by an IPCE test system fabricated by Newport Co.

## 3. Results and discussion

Figure 1 shows the deposition process of c-TiO$_2$ layer by PLD. With high-powered nanosecond laser focused on TiO$_2$ target, high-temperature and high-density plasma (Ti$^{4+}$ and O$^{2-}$) is produced and deposit on FTO glass.[24] As shown in the inset image in Figure 1, FTO glass selected as the substrate is fairly transparent before deposition and became a bit black after depositing 32 nm thick c-TiO$_2$ layer. Subsequently, the c-TiO$_2$ thin film prepared by PLD got transparent after annealing in air. EDS spectra of the thin film before and after annealing, shown in Table S1, reveal that the atomic ratio of Ti/O significantly decreases from 0.7 to 0.57 after annealing, indicating the evident existence of oxygen vacancies in the as-prepared thin films and effective remove of oxygen vacancies after annealing in air for enough time.

Subsequently, we measured the light absorbance spectrum of bare FTO glass, 32 nm thick pre-annealed and post-annealed samples deposited on FTO glass which is displayed in Figure S1(a). It is obvious that the post-annealed sample whose absorbance coefficient is lower than 0.4, which could transmit more light than the pre-annealed one with the absorbance coefficient approaching 0.5. Thus, more oxygen vacancy could lead in a non-transparent color of c-$TiO_2$ layers so that less light can be absorbed by perovskite layers in devices. What's more, according to Figure S1(b), we can see that the intensity of PL peak for post-annealed c-$TiO_2$ layers is obviously lower than that of peak for pre-annealed c-$TiO_2$ layers. This indicates that oxygen vacancy is disadvantageous for electron extraction and transport from perovskite to the ETL, resulting in higher charge carrier recombination.[25] In addition, the intensity of PL peak for post-annealed c-$TiO_2$ layers prepared by PLD is larger than that for c-$TiO_2$ layers prepared by spin coating post-annealed c-$TiO_2$ layers, which demonstrates that more compact ETL can contribute to decreasing of charge carrier recombination and mobility of charge carrier into the FTO layer. Figure S2 shows the optical transmission spectrum of FTO glass and the 32 nm thick post-annealed c-$TiO_2$ layer on FTO glass. Referring to this spectrum, we can find that FTO glass as the substrate shows transmittance from 60 % to 80% in the wavelength range 400–800 nm and the value of transmittance for the sample is around 50% – 70% in the visible region and sharply decreases to 0 % in the ultraviolet range in accordance with the characteristic of c-$TiO_2$ layers.[26,27] The relationship between the thickness (y) of c-$TiO_2$ thin films and pulse number (x) can be linearly fitted using an equation: y (nm)

=0.00843x, which is displayed in Figure S3.

Spin coating is a common method widely used in preparation of c-TiO$_2$ thin films.[28,29]. We deposited a 32 nm thick c-TiO$_2$ thin films (4000 pulse) on FTO glass by PLD (the device shows the best performance). In comparison, another TiO$_2$ thin film was prepared by spin coating whose optimum thickness in our preparation was nearly 100 nm. Figure 2 shows polarizing microscope plane-view images of FTO glass, c-TiO$_2$ layer prepared by PLD and TiO$_2$ layer prepared by spin coating. We can find that the c-TiO$_2$ layer prepared by PLD shown in Figure 2(b) are similar with FTO glass shown in Figure 2(a). However, in comparison with the former, the c-TiO$_2$ layer prepared by spin coating shown in Figure 2(c) obviously appears distribution of different colors in the c-TiO$_2$ thin film surface, indicating that c-TiO$_2$ thin films on spin coating are not uniform compared with those on PLD. From Figure 3(a)-(c), we can easily find that the 32 nm thick c-TiO$_2$ layer prepared by PLD (Figure 3(b)) is not visible compared with the spin coated one. The thickness of the former is much smaller than that of the latter. As shown in Figure 3(d), bare FTO grains with irregular shape of FTO glass could be clearly recognized. From Figure 3(e), we could see the surface morphology of the FTO glass overlaid with a 32 nm thick TiO$_2$ layer still remain legible. In comparison, the original surface morphology of FTO glass with a spin coated TiO$_2$ layer is not observable, as shown in Figure 3(f). Figure 4(a)-(c) display AFM images of surface, the root-mean-square (RMS) surface roughness of the 4000-pulse TiO$_2$ thin film is 33.7±5.835 nm which is close to that (35.28±9.717 nm) of FTO glass and nearly twice larger than that of the spin coated TiO$_2$ thin film (15.97

±2.275 nm).

As is shown in Figure 5(a), we use a method to measure compactness of 60 nm thick c-TiO$_2$ layers on spin coating and PLD by evaporating Ag electrodes on c-TiO$_2$ layers and obtaining resistance values between two closed Ag electrodes.[30] Comparing these resistance values can make us clearly know distribution of pinholes density which reflects compactness of c-TiO$_2$ layers. A more compact c-TiO$_2$ layers which has lower pinhole density can effectively increase the resistance value. Seen from Figure 5(b), we can know that all resistance values of c-TiO$_2$ layers prepared by spin coating are $10^1 \sim 10^2$ Ω much lower than those of c-TiO$_2$ layers prepared by PLD which are $10^3 \sim 10^4$ Ω. So, c-TiO$_2$ layers prepared by PLD are more compacter than c-TiO$_2$ layers prepared by spin coating.

We further study the application of c-TiO$_2$ thin film prepared by PLD in the structure of PVSCs as the ETL, shown in Figure 6(a). Figure 6(b) shows the plane-view SEM image of the MAPbI$_3$ thin film with grain size of around 1 μm as the light absorption layer in devices. The XRD pattern of MAPbI$_3$ thin film fabricated in our work is shown in Figure 6(c), indicating the pure phase without any impurities. Figure 6(d) displays the light absorption spectrum and the steady-state PL spectrum of MAPbI$_3$ thin film which are in accordance with the pure MAPbI$_3$.[31] The transient-state PL spectrum shown in Figure 6(e) is fitted with a bi-exponential equation: $I(t) = I_0 + A_1 \exp(-\frac{t-t_0}{\tau_1}) + A_2 \exp(-\frac{t-t_0}{\tau_2})$ ($A$ is the decay amplitude, $\tau$ is the decay time and $I_0$ is a constant) and the average decay time ($\tau_{avg.}$ = 21.59 ns) can be obtained by the equation: $\tau_{avg.} = \frac{\sum A_i \tau_i^2}{\sum A_i \tau_i}$.[32,33] The fitting results of the transient-state PL spectrum displayed in Table S2. The cross-sectional SEM image of PVSCs fabricated in our work is exhibited in Figure S4. The device is comprised of a 100 nm thick Ag electrode, a 200 nm thick HTL, a 250 nm thick $MAPbI_3$ layer, a 160 nm thick m-$TiO_2$ layer, a c-$TiO_2$ layer of various thickness (4 nm to 40 nm) on a FTO glass substrates.

As shown in Figure 7(a), it is obvious that the PCE of devices gradually increases when the thickness of the c-$TiO_2$ layer increases to 32 nm and then decreases with further increasing thickness above 32 nm. The range of PCE of devices with 32 nm thick c-$TiO_2$ layers is from 11.28% to 13.95%, which is higher than that of the others. Figure 7(b) shows that FF of devices with 4 nm thick c-$TiO_2$ layers is below 40%, and continues to increase with increasing $TiO_2$ thickness, and nearly saturates at around 65% with thickness above 25 nm. Figure 7(c) exhibits that ranges of $V_{oc}$ of all devices are all from 0.9-1.0 V without significant dependence on the thickness of the c-$TiO_2$ layer. Figure 7(d) shows that the relationship between $J_{sc}$ of devices and the thickness of the c-$TiO_2$ layer. It can be seen that $J_{sc}$ of devices increases with increasing c-$TiO_2$ thickness to 16 nm, and then keeps nearly constant, and finally decreases with further increasing thickness above 32 nm. Therefore, the increasing of FF and $J_{sc}$ cooperatively contributes to the increase of PCE and best performance obtained is obtained with c-$TiO_2$ layer thickness of 32 nm. The J-V curve and IPCE spectrum of the best PVSCs with 32 nm thick c-$TiO_2$ layer are shown in Figure 8(a)-(b). The $J_{sc}$ is 21.83 mA/$cm^2$, the $V_{oc}$ is 1.0 V, FF is 63.98% and the PCE of the solar cell is 13.95%.

As shown in Figure S6, an as-prepared 400 nm thick $TiO_2$ film is black while becomes transparent after annealing it at 500 °C in air for enough time, indicating the high concentration of oxygen vacancies in thick as-prepared $TiO_2$ layer. This is due to that increased thickness is disadvantageous for the removal of internal oxygen vacancies too far away from surface of thick as-prepared $TiO_2$ film. Oxygen vacancy related defects are adverse for transport of charge carrier, leading to the evident decrease of $J_{sc}$.[34] Remnant of oxygen vacancy can cause worse light transmitting mentioned previously. In addition, increasing thickness of $TiO_2$ film also slightly add series resistance of devices which can reduce $J_{sc}$ of devices.[23] An appropriate thickness for c-$TiO_2$ layers prepared by PLD can allow us to obtain a best performance of devices and the optimum thickness is around 32 nm. It should be noted that a smoother substrate compared with FTO can result in a smaller optimum thickness than 32 nm.[35] The ultrathin $TiO_2$ layer is also beneficial for light transmitting to $CH_3NH_3PbI_3$ layer, causing more light to be absorbed and transform into electrons and decreasing series resistance which can obviously improve performance of devices, especially in planar-structured solar cells.[36]

Conclusion

In summary, c-$TiO_2$ layers with various thicknesses were prepared by PLD, and applied as ETL in the $MAPbI_3$-based solar cells. $TiO_2$ layers prepared by PLD can be more uniform and compact, which may contribute to produce large area PVSCs. By analyzing the performance and EIS spectra of devices, the optimum thickness of c-$TiO_2$ layers has been determined to be around 32 nm which is much thinner than those prepared by solution deposition method. c-$TiO_2$ layer with optimized thickness of 32 nm has been applied in the CH$_3$NH$_3$PbI$_3$-based solar cell devices and the highest efficiency of 13.95 % with Jsc of 21.83 mA/cm$^2$, V$_{oc}$ of 1.0 V and FF of 63.98% has been achieved. Importantly, the introduction of PLD into PVSCs can make ion doping in c-TiO$_2$ layers easy and convenient which can accelerate the development of the high-quality ETL to further improve the photovoltaic performance and the industrial application of PVSCs.

## Acknowledgments

This work is supported by the National Natural Science Foundation of China (51471085, 51771053), the Natural Science Foundation of Jiangsu Province of China (BK20151400), and the open research fund of Key Laboratory of MEMS of Ministry of Education, Southeast University.

## Additional Information

**Supplementary information**

**Competing interests:** The authors declare no competing interests.

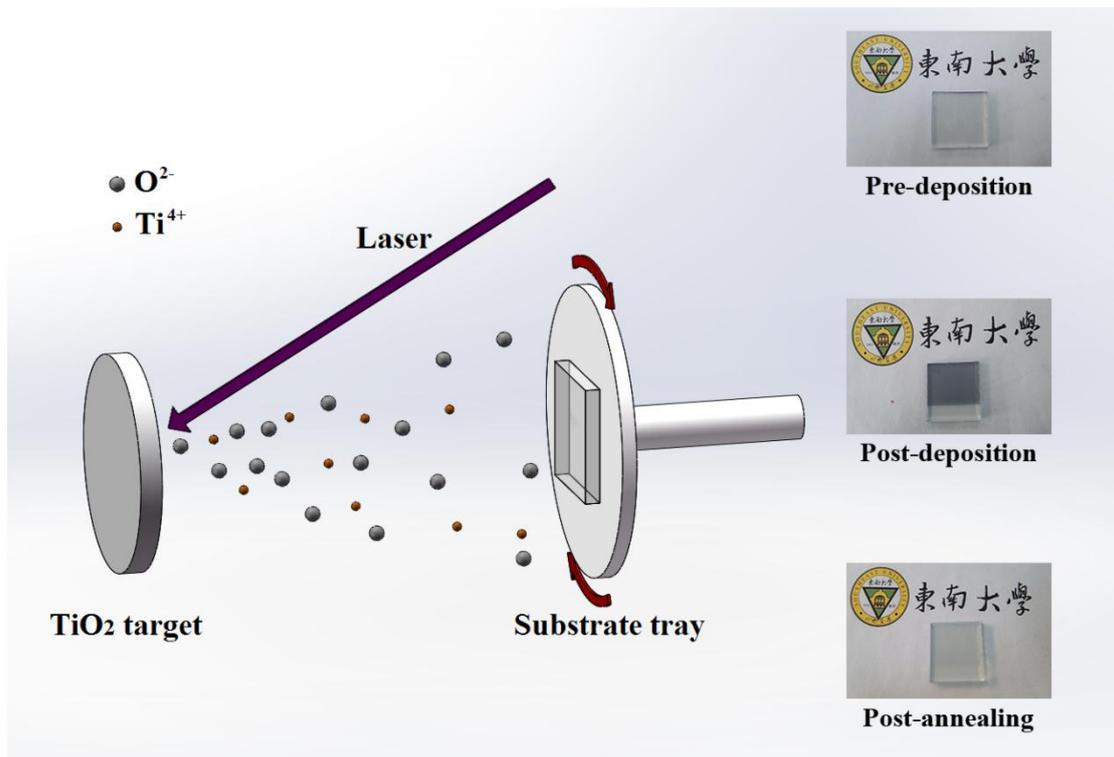

**Figure 1.** Schematic representation concerning the preparation of the compact TiO$_2$ layer by PLD. Inset shows the photos of 32 nm thick TiO$_2$ films.

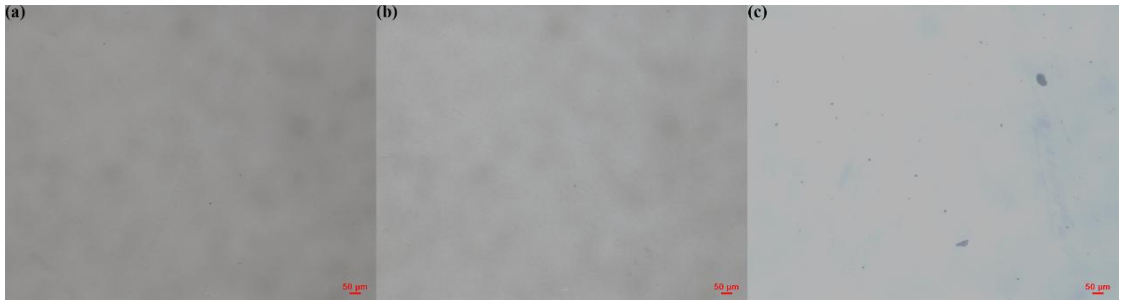

**Figure 2.** Polarizing microscope plane-view images of (a) FTO glass, (b) 30 nm thick compact TiO$_2$ layer prepared by PLD and (c) TiO$_2$ layer prepared by spin coating.

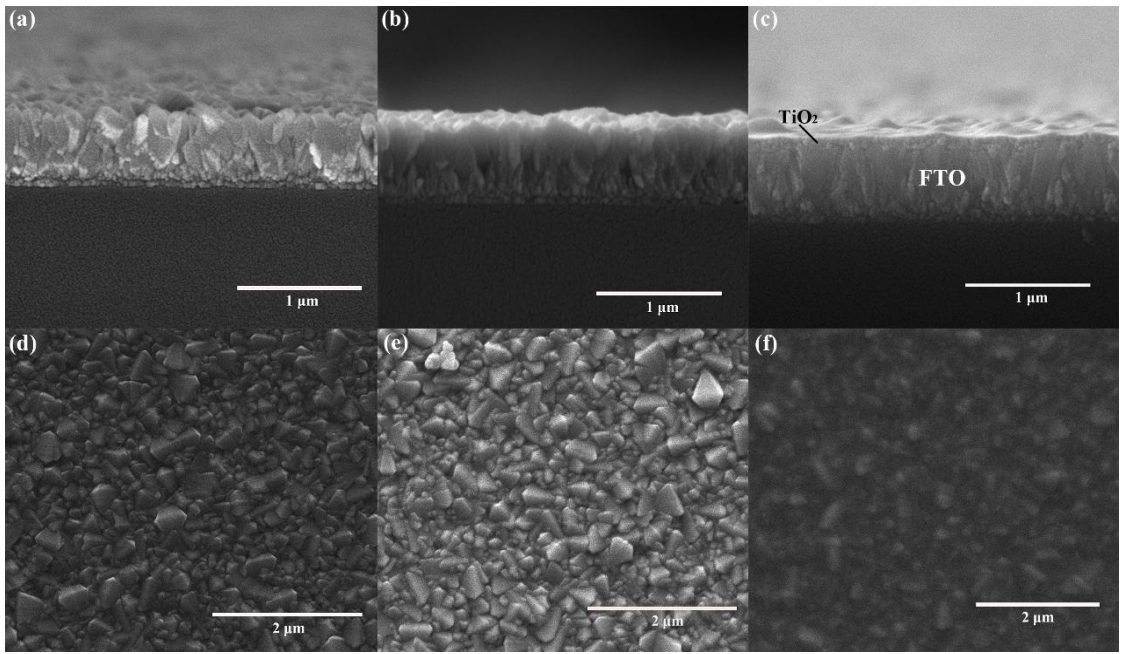

**Figure 3.** Cross-sectional SEM images of (a) FTO glass, (b) 32 nm thick compact TiO$_2$ layer prepared by PLD and (c) TiO$_2$ layer prepared by spin coating. (d) to (f) are plane view SEM images, corresponding to (a) to (c), respectively.

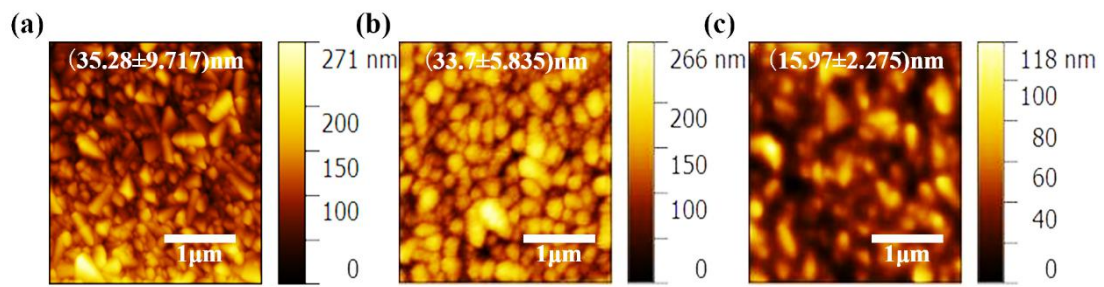

**Figure 4.** AFM images of (a) FTO glass, (b) 32 nm thick compact $TiO_2$ layer prepared by PLD and (c) compact $TiO_2$ layer prepared by spin coating with their RMS value of surface roughness.

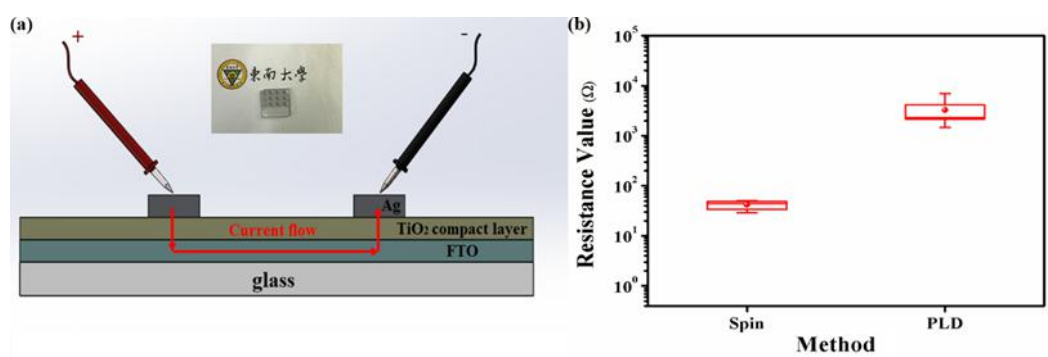

**Figure 5.** (a) The schematic of the resistance measurement method. (b) Resistance values range of c-TiO$_2$ layers prepared by spin coating and PLD.

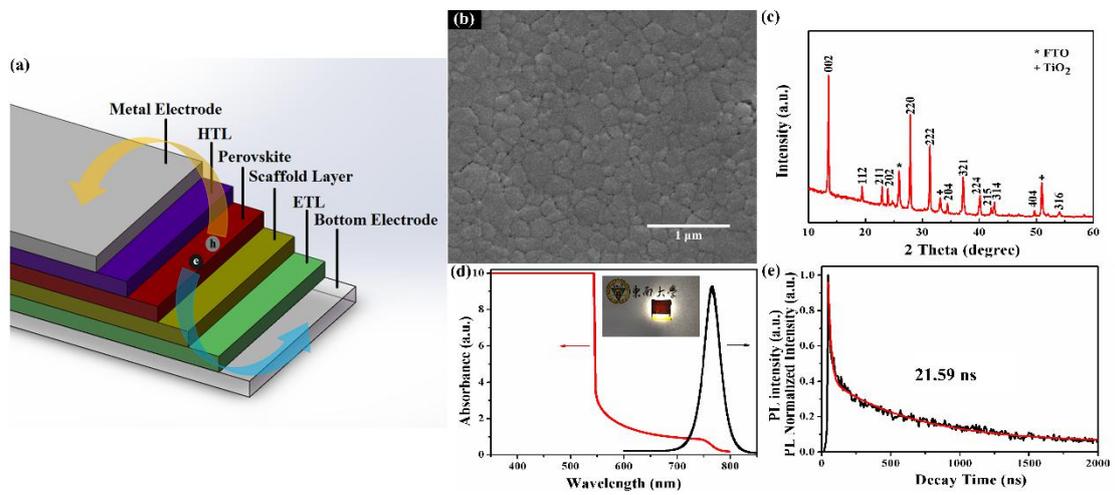

**Figure 6.** (a) The diagram of perovskite solar cells fabricated in this work. (b) The plane-view SEM image, (c) XRD pattern, (d) light absorption spectrum and steady-state PL spectrum and (e) transient-state PL spectrum of the MAPbI$_3$ thin film.

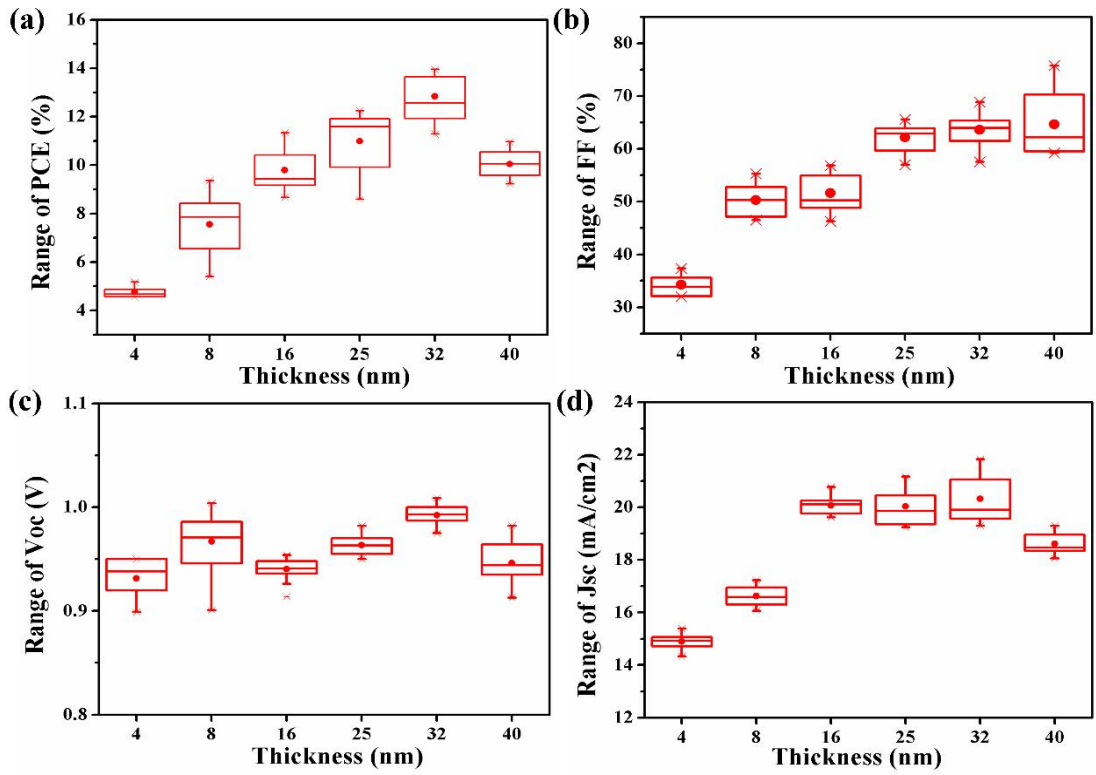

**Figure 7.** (a) PCE, (b) FF, (c) $V_{OC}$ and (d) $J_{SC}$ of devices with compact $TiO_2$ layers of various thicknesses ranging from 4 nm to 40 nm.

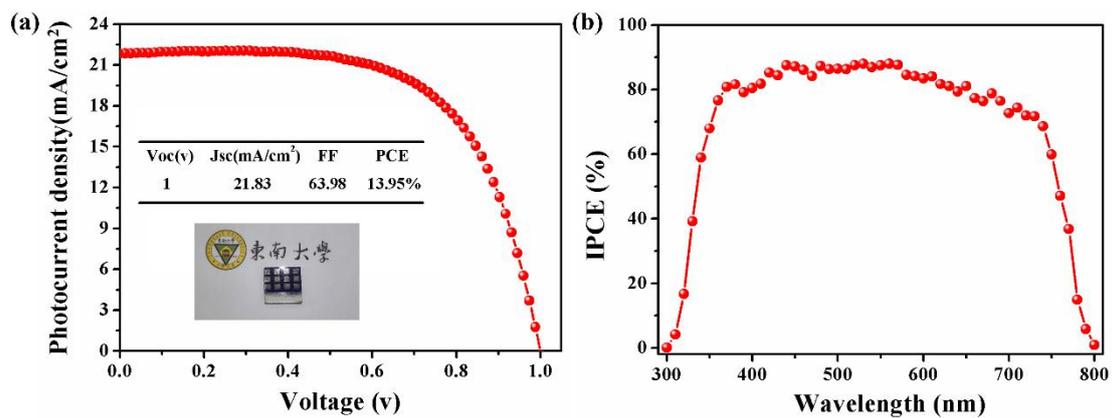

**Figure 8.** (a) J-V curve and (b) IPCE of the CH$_3$NH$_3$PbI$_3$-base solar cell with the best performance.